\documentstyle[aps,psfig,amssymb]{revtex}
\begin{document}
\draft
\title{Simulation and experimental investigation of cellular material breakage
using the pulsed electric field treatment}
\author{
N. I. Lebovka$^{a,b}$
\footnote{E-mail:Nikolai.Lebovka@utc.fr}
,
M.I. Bazhal$^{a,c}$
\footnote{E-mail:Maksym.Bazhal@utc.fr}
, \&
E. I. Vorobiev$^{a}$
\footnote{E-mail:Eugene.Vorobiev@utc.fr}
}

\address{
$^{a}$D\'{e}partement de G\'{e}nie Chimique, Universit\'{e} de
Technologie
de Compi\`{e}gne, Centre de Recherche de Royallieu, B.P. 20529-60205
Compi\`{e}gne Cedex,
France\\
$^{b}$Institute of Biocolloidal Chemistry named after F.D. Ovcharenko, NAS
of Ukraine, 42, blvr.Vernadskogo, Kyiv, 252142, Ukraine\\
$^{c}$Ukrainian State University of Food Technologies, 68, Volodymyrska
str., Kyiv, 252033, Ukraine\\
}

\date{June 9, 1999}

\maketitle

\begin{abstract}
We consider the simplified dielectric breakage model used for simulation of
the kinetics of cellular material breakage under the
pulsed electric field (PEF) treatment. The model is based on an effective
media approximation, which
includes equations with the same morphology parameters as in percolation
theory. The probability of a whole cell breakage by the pulse with $t_{i}$
duration is estimated on the basis of electroporation theory. We account for
the bridging effect resulting from the deviations of the local conductivity
near the selected cell from the average effective media conductivity.
The most important feature of the proposed model is the existence of the
``jamming'' behaviour occurring sometimes in experimental observations of
the biological tissue breakage. The different transitions corresponding to
the ``jamming'' steps are identified. The experimental results are obtained
for thin apple slices treated with electric pulses at field
strengths $E=0.2-2.2$ kV cm$^{-1}$, pulse durations $t_{i}=10-100$ $\mu$s, pulse
repetition times $t=10-100$ ms and the number of pulses $N=1-100000$. The
model gives results consistent in general with the experimental
observations. We discuss the correlation between the degree of cellular material
destruction, field strength, time of PEF treatment and power consumption.
\end{abstract}

\pacs{
}

\bigskip\bigskip


\begin{tabular}{ll}
{\bf Notation} &\\
C$_{m}$ & specific capacity of membrane, F m$^{-2}$ \\
C$^{\ast }$ & = C$_{m}$($\varepsilon _{w}$/$\varepsilon _{m}$-1)/(2$\gamma $)
\\
C & specific heat of cellular material, J kg$^{-1}$ K$^{-1}$ \\
$d_{m}$ & membrane thickness, m \\
$d_{c}$ & cell diameter, m \\
$E$ & electric field strength, kV cm$^{-1}$ \\
$\Delta F^{\ast }$ & $=\pi \omega ^{2}/(kT\gamma )$, reduced critical free
energy of pore formation \\
$g$ & breakage probability for a membrane\\
$j$ & current density, A m$^{-2}$ \\
$k$ & Boltzmann constant, 1.38065812x10$^{-23}$J K$^{-1}$ \\
$L$ & $=d_{c}N_{z}$, total width of a sample\\
$m$ & $=d_{c}/d_{m}$ \\
$n$ & number of intact cells neighbouring of a given cell \\
$N$ & number of pulses \\
$N_{b}$ & number of pulses at the moment of the total dielectric breakdown
of material \\
$N_{m}$ & $\,$number of intact cells \\
$N_{x}$,$N_{y}$,$N_{z}$ & dimensions of a lattice \\
$P$ & degree of biological tissue destruction \\
$P_{n}$ & local degree of cell destruction \\
$P_{p}$ & critical percolation degree of destruction \\
$r_{m,c,i}$ & = $d_{m,c,c}/(S_{m}\sigma _{m,c,i})$, resistance, Ohm \\
$R$ & total resistance of a linear chain, Ohm \\
$S_{m}$ & $\sim d_{c}^{2}$, cross section area of a cell or a membrane, m$%
^{2}$ \\
\end{tabular}

\begin{tabular}{ll}
$t$ & scaling exponent \\

$t_{i}$ & pulse duration, $\mu $s \\
$\Delta t$ & pulse repetition time, ms \\
$T$ & temperature, K \\
$\Delta T$ & temperature increase over initial value, ${{}^{\circ }}$C \\
$u$ & transmembrane voltage, V \\
$u_{o}$ & midpoint of a probability transition function $g$, V \\
$\Delta u$ & width of a probability transition function $g$, V \\
$U$ & external voltage, V \\
$W$ & moisture content, \% \\
\end{tabular}

\bigskip
{\it Greek letters}

\begin{tabular}{ll}
$\alpha $ & geometry factor \\
$\gamma $ & surface tension of membrane, N m$^{-1}$ \\
$\varepsilon _{w}$ & $=80$, dielectric constant of water \\
$\varepsilon _{m}$ & $=2$, dielectric constant of membrane \\
$\varphi $ & $=P/(1-P)$ \\
$\lambda $ & $=d_{c}\sigma _{m}/d_{m}\sigma _{c}$ \\
$\rho $ & density of cellular material, kg m$^{-3}$ \\
$\sigma $ & conductivity, Sm$^{-1}$ \\
$\sigma _{i}$ & $=\sigma _{c}\lambda /(1+\lambda )$, conductivity of an
intact cell \\
$\Sigma $ & $=\Sigma _{e}/\Sigma _{c}$ \\
$\Sigma _{o}$ & $=\Sigma _{i}/\Sigma _{c}$ \\
$\Sigma _{e,c,i}$ & $=\sigma _{e,c,i}^{1/t}$ \\
$\tau $ & lifetime of a membrane\\
$\tau _{\infty }$ & parameter, lifetime of a membrane at $T=\infty$\\
$\psi $ & $=1/P_{p}-1$\\
$\omega $ & linear tension of membrane, N 
\end{tabular}

\bigskip
{\it Subscripts}

\begin{tabular}{ll}
$a$ & after treatment \\
$b$ & before treatment \\
$c$ & cell \\
$e$ & effective \\
$i$ & intact cell \\
$j$ & juice \\
$m$ & membrane \\
$t$ & total
\end{tabular}

\bigskip
{\it Abbreviations}

\begin{tabular}{ll}
PEF & pulsed electric field \\
LC &  linear chain \\
MF & mean field \\
RSA & random sequential adsorption \\
RSO & random sequential occupation \\
\end{tabular}

\section{Introduction}

Pulsed electric field (PEF) treatment is a new and promising nonthermal
processing method for heterogeneous cellular materials. PEF methods are based
on the effect of cell transformation or rupture under an external electric
field, which results in increase of the electric conductivity and
permeability of cellular material. This effect known as dielectric breakdown
(Zimmermann, Pilwat \& Riemann, 1974), or electroplasmolysis (Scheglov,
1983), can be explained generally by electroporation, i.e., electric field
induced formation and growth of pores in biological membranes resulting from
their polarization. The method of electroporation became very popular
because it was found to be an exceptionally practical way of transferring
drugs, genetic material (e.g. DNA), or other molecules inside the cells
(Chang, Chassy, Saunders \& Sowers, 1992). Recently, scientists began to use
the PEF methods for treatment of liquid food (fruit juices, milk etc.), as
an alternative to high temperature preservation (Barbosa-Canovas,
Pothakamury, Palou \& Swanson, 1998; Hulsheger, Potel \& Niemann, 1983).
PEF protocols also have been recently introduced for cellular tissue
treatment (Knorr, Geulen, Grahl \& Sitzmann, 1994; Knorr, 1997).

However, presently the wide spreading of PEF treatment for the purposes of
nonthermal processing of heterogeneous cellular materials is restrained by the
poor reproducibility of experimental data, the unclear mechanism of
dielectric breakdown in the cellular systems, and the absence of criteria
for choosing optimal parameters of HEFP treatment. The most optimal for the
tasks of this kind is to apply computer simulation methods. There exist many
types of models for simulation of dynamic breakage processes in random
disordered systems. The most popular is the model of random resistor (or
random conductor) networks (Arcangelis, Redner \& Herrmann, 1985; Duxbury \&
Beale, 1995). The models of this type are applied for simulation of the
physical behavior of granular superconductor, metal-insulator composites and
various other disordered materials. Usually, the breakdown process kinetics
is investigated at the network models of different types through the
iterative solving of the Ohm-Kirchhoff's system of equations with allowance
for boundary conditions and with choosing the relevant initial conditions
for breakdown initialization (Lebovka \& Mank, 1992). But all the above
mentioned models are highly time consuming and they require a lot of
CPU-time for the real cellular material simulation (Press, Teukolsky,
Vetterling \& Flannery, 1997).

The objective of this study is to develop the simulation model describing
the breakage kinetics under conditions of PEF treatment and to correlate
experimental data for heterogeneous cellular materials with simulation results.
We consider the very simplified model of cellular material breakage under the
PEF, which is based on a generalised approximation of effective medium.
This model takes into account the percolation properties of medium and local
bridging effects. This approach is, certainly, rather rough, same as all the
other effective medium approaches, but it allows us to obtain the realistic
description of breakage processes and trends of kinetics.

\section{Background Theory and Simulation Model}

\subsection{Field Induced on a Cell Membrane in an External Field}

We consider at first the case of an individual cell inside the homogeneous
medium having a low effective conductivity $\sigma _{e}$ placed into an
external electric field of strength $E$. The maximum voltage $u$ between the
internal and external cellular surfaces of membrane (or transmembrane
voltage) induced on cell poles by the external field, is equal to (Neumann,
Sprafke, \& Boldt, 1992):
\begin{equation}
u=\alpha fd_{c}E  \label{e01}
\end{equation}
where $\alpha $ is the geometry factor equal to 0.75 for the spherical
geometry of cell and to 1 for a cell of cubic geometry, and $f$ depends on the
electrical and geometrical properties of the cell (Kotnik,
Miklavcic \& Slivnik, 1998)
\begin{equation}
f=\frac{\sigma _{e}(6m\sigma _{e}-(12m^{2}-8m^{3})(\sigma _{m}-\sigma _{c}))%
}{(\sigma _{m}+2\sigma _{e})(\sigma _{m}+\sigma _{c}/2)-(1-2m)^{3}(\sigma
_{e}-\sigma _{m})(\sigma _{c}-\sigma _{m})}.  \label{e02}
\end{equation}

Taking into account the low conductivity of membrane,
$\sigma _{m}/\sigma _{c}<$$<$1 and expanding
Eq. (\ref{e02}) in power of $\sigma _{m}/\sigma _{c}$, we obtain:
\begin{equation}
f=(1+\frac{\sigma _{m}}{2\sigma _{e}K})/K+O\left[ \left( \sigma _{m}/\sigma
_{c}\right) ^{2}\right] ,  \label{e03}
\end{equation}
where $K=1+\lambda (2+\sigma _{c}/\sigma _{e})/4$ , and under the usual
conditions (standard values of $d_{c}$, $d_{m}$, $\sigma _{m}$, $\sigma _{c}$
and $\sigma _{e}$ parameters for this problem are collected by Kotnik,
Miklavcic and Slivnik (1998)) $\lambda \sim 10^{-1}-10^{-2}<$$<1$ and $f\sim
1$. For the purposes of estimation we shall consider further on only the cells
of cubic geometry, so we put $\alpha =1$. In this approximation, $f$ is equal
to the normalised transmembrane voltage $u/d_{c}E$.

Now we consider a problem of transmembrane voltage $u$ estimation for a
cell inside a biological tissue. In this case, the external media is not
homogeneous, so, Eqs. (\ref{e01}) and (\ref{e02}) are inaccurate. They can be
used only as approximate estimation by assuming a certain effective
value of $\sigma _{e}$. This is just the mean field (MF) type of
approximation, which does not take local conductivity fluctuations into
account. It can be assumed, as a rough approximation, that in an inhomogeneous
medium the equation (\ref{e01}) may also be used for determination of the $u$
values, however, $f=f(r)$ will depend on the space co-ordinate $r$ and will
vary from cell to cell.

As it follows from Eq. (\ref{e03}), the value of f gradually increases with
the increase of external medium effective conductivity $\sigma _{e}$, which
depends on the biological tissue destruction degree, $P$. Here we define the
biological tissue destruction degree as:

\begin{equation}
P=1-N_{m}/N_{t}.  \label{e04}
\end{equation}

As a more precise estimation of $\sigma _{e}$ versus $P$
dependence, we have used the generalised effective media equation with the
same morphology parameters as in percolation theory (McLachlan, 1989):

\begin{equation}
\frac{\varphi (1-\Sigma )}{1+\psi \Sigma }+\frac{\Sigma _{o}-\Sigma }{\Sigma
_{o}+\psi \Sigma }=0.  \label{e05}
\end{equation}

This equation reduces to the well-known Bruggeman's symmetric equation when
$t=1$ (Landauer, 1978). Everywhere throughout this work we use the values of
$t=2.3$, $P_{p}=0.307$, which are typical for the 3d simple cubic lattice (Sahimi, 1993).
In the general case we obtain the following solution of Eq. (\ref{e05}):

\begin{equation}
\sigma _{e}(P)=\sigma _{c}\left[ \left(
\begin{array}{l}
\varphi (\psi -\Sigma _{O})+\Sigma _{O}\psi -1+ \\
+\sqrt{\varphi ^{2}(\psi +\Sigma _{O})^{2}+2\varphi \lbrack \Sigma _{O}(\psi
+1)^{2}-\psi (\Sigma _{O}-1)^{2}]+(\psi \Sigma _{O}+1)^{2}}
\end{array}
\right) /\left( 2\psi (\varphi +1)\right) \right] ^{t}.  \label{e06}
\end{equation}
which can be easily used for the numerical estimation of $\sigma _{e}$
versus $P$ dependency.

From Eq. (\ref{e06}) we can easily obtain $\sigma _{e}=\sigma _{i}=\sigma
_{c}\lambda /(1+\lambda )$ at $P=0$ and $\sigma _{e}$=$\sigma _{c}$ at $P=1$.

The example of $f$ versus $P$ dependence on the mean field approximation
obtained with the help of Eqs. (\ref{e02}) and (\ref{e06}) is presented in
Fig. \ref{f1} (line 1). At low degrees of the biological tissue
destruction, $P\rightarrow 0$, the effective conductivity approximately
equals to $\sigma _{e}=\sigma _{i}=\sigma _{c}\lambda /(1+\lambda )$. At
complete destruction of a biological tissue, $P\rightarrow 1$, we have $%
\sigma _{e}\approx \sigma _{c}$. Thus, we obtain the following limiting
relations in MF approximation:

\begin{eqnarray}
f(P &\rightarrow &0)=1/(2+3\lambda )\approx 1/2,\text{and}  \label{e07} \\
f(P &\rightarrow &1)=1/(1+3\lambda )\approx 1  \nonumber
\end{eqnarray}
i.e., the value of $f$ approximately doubles with the biological tissue
destruction increase within the range of $P=0-1$.

In fact, the change in $f$ with increasing biological tissue
destruction degree ($P$) is impossible to describe within the frames of such
simplified approach, because of substantial local fluctuations of the
external medium conductivity existing in the real inhomogeneous biological
tissues. Such fluctuations are especially high near the punctured cells,
where the local values of conductivity can exceed the effective average
value $\sigma _{e}$ by some orders of magnitude. The MF approximation works
especially poor at low degrees of biological tissue destruction, $%
P\rightarrow 0$, when the local conductivity fluctuations are high. In order to
demonstrate the complex behaviour of $f$ versus $P$ dependence in a real
biological tissue, we have considered the oversimplified linear chains (LC)
approximation as a simulation of a real tissue structure. The tissue is
simulated by linear chains consisting of $N_{t}$ cells of a cubic-like
geometry ($\alpha =1$). The total resistance of such linear chain consisting
of serially connected membrane resistances ($r_{m}=d_{m}/(\sigma _{m}S_{m})$) and
intercellular fluid resistances ($r_{c}=d_{c}/(\sigma _{c{}}S_{m})$) can be
estimated using the following relation:
\begin{equation}
R=(N_{t}r_{c}+N_{m}r_{m}).  \label{e08}
\end{equation}

The density of current running through such linear chain is equal
to:
\begin{equation}
j=U/(RS_{m})=U/(N_{t}d_{c{}}/\sigma _{c{}}+N_{m}d_{m}/\sigma _{m})=\sigma
_{e}U/(N_{t}d_{c})=\sigma _{e}E,  \label{e09}
\end{equation}
where
\begin{equation}
\sigma _{e}=\lambda \sigma _{c}/(\lambda +1-P),  \label{e10}
\end{equation}
is an effective chain conductivity, and $E = U/L = U/(N_{t}d_{c})$.

The percolation point is observed in LC-approximation at $P=P_{p}=1$, and it
can be easily shown, that the more general Eq. (\ref{e06}) reduces to the
Eq. (\ref{e10}) in the case of $t=1$.

The voltage drop at a single membrane is equal to:

\begin{equation}
u=jS_{m}r_{m}=jd_{m}/\sigma _{m}=d_{c}E\sigma _{e}/(\lambda \sigma _{c}),  \label{e11}
\end{equation}
or we obtain for $f=u/d _{c}E$ (at $\alpha =1$):
\begin{equation}
f=\sigma _{e}(P)/\lambda \sigma _{c}.  \label{e12}
\end{equation}

The example of $f$ versus $P$ dependence for the linear chain approximation
obtained with the help of Eqs. (\ref{e06}) and (\ref{e12}) is presented in
the Fig. \ref{f1} (line 2).

We obtain from Eqs. (\ref{e06}) and (\ref{e12}) in LC approximation with
account for $\lambda $ $<$$<1$:

\begin{eqnarray}
f(P &\rightarrow &0)=1/(1+\lambda )\approx 1,\text{ \ and}  \label{e13} \\
f(P &\rightarrow &1)=1/\lambda >>1  \nonumber
\end{eqnarray}
i.e. in this case the value of f can considerably increase in the course of
biological tissue destruction, and it is in contradiction with the
conclusion, which follows from Eq. (\ref{e07}) on the basis of MF
approximation.

Moreover, MF approximation gives us $f(P\rightarrow 1)\approx 1$, but in LF
approximation we have $f(P\rightarrow 0)=1/(1+\lambda )\approx 1$. How we
can clear this contradiction? MF approximation is good only at large values
of $P\rightarrow 1$, when all the cells of tissue are destroyed and the
media is practically homogeneous. On the other side, LC approximation works
well only at small values of $P\rightarrow 0$. In this latter case we can
neglect any bridging effects, which are very important in the middle of
the range of $P$ values. The bridging effect results from the deviations of
local conductivity near the selected cell from the average
effective conductivity of the whole media. Analysing the parallel
electrical circuits in the near-neighbours environments of a given cell, we
can easily approximate the averaged local cell resistance $r_{n}$ as

\begin{equation}
r _{n}\approx 1/((1-P_{n})/r _{i}+P_{n}/r _{c}).  \label{e14}
\end{equation}

For the cells arranged in the sites of a simple cubic lattice we have $%
P_{n}=(1-n/6)$, where $n$ is the discrete number of intact cells
neighbouring a given cell. Then the voltage drop at a single membrane
equals to:

\begin{equation}
u=jr _{n}S_{m}r_{m}/r_{i}=\frac{d_{c}E\sigma _{e}}{\lambda \sigma _{c}(1+P_{n}/\lambda)},  \label{e15a}
\end{equation}
and  we obtain the following generalised equation for estimation of the local
value $f_{n}=u/d _{c}E$

\begin{equation}
f_{n}(P,P_{n})=\sigma _{e}(P)/(\lambda +P_{n})\sigma _{c}  \label{e15}
\end{equation}
where in the general case the value of $\sigma _{e}$ at given $P$ may be
determined from Eq. (\ref{e06}).

Figure \ref{f2} shows the example of local conductivity $f_{n}$ versus $P$
dependencies obtained at various $n$ with the help of Eqs. (\ref{e06}) and (%
\ref{e15}). We see that the local behaviour of $f_{n}(P,P_{n})$ function can
be rather complex and it can reflect the local conductivity changes near the
chosen cell. We have shown with the dashed curve the imaginary fluctuations
of $f_{n}$ with $P$ increase and successive breakdown of neighbouring cells.
We understand that this approach proposed for estimation of the $%
f_{n}(P,P_{n})$ behaviour is very simplified. Particularly, here we
overestimate the importance of the lattice contribution to the $%
f_{n}(P,P_{n})$ behaviour due to the discreteness of this problem, and in
the case of continuum more smooth behaviour of $f_{n}(P,P_{n})$ can be
observed for the continual parameter $P_{n}$.

The behaviour of the averaged value $f(P)=<f_{n}(P,P_{n})>$ (here $<$$\ldots
$ $>$ implies averaging over all the cells of a system) will depend on the
spatial distribution of broken cells in such system. Let us consider as an
example, the $f(P)$ behaviour for the model of random sequential occupation
(RSO) of the lattice sites with the broken cells. This problem is very
complex and has no exact solution. Here, the Monte Carlo simulation is a
rather simple and useful method allowing to get the $f(P)$ function for the
given type of cell distribution in the system. We have considered the cells
located in the sites of a simple cubic lattice with the following
dimensions: $100$x$100$x$100$. The simulation consisted of successive random
choice of an intact cell, its further destruction and averaging of the
calculated $f$-values over the whole lattice. We used the periodical
boundary conditions along all the $x$, $y$ and $z$ directions in order to
reduce the influence of boundaries. The results were averaged over 10
different initial configurations. The example of $f$ versus $P$ dependence
obtained as described above is presented in the Fig. \ref{f1} (line 3). The
dashed line 4 corresponds to $f$ dispersion ($\Delta f$) and characterises
fluctuations of $f$ in the system. We see that on increase of the
destruction degree ($P$) the values of $f$ first decrease, then pass through
a minimum and increase again. The point of minimum $f(P)$ approximately
corresponds to the maximal fluctuations of the local values $f_{n}$ in a
system, i.e., to the maximum of $\Delta f$. The important positive feature of this
method of $f(P)$ estimation is as follows: we get $f\approx 1$ in both limit
cases of $P\rightarrow 0$ and $P\rightarrow 1$, i.e., we remove here the
contradiction between MF and LC approximations.

\subsection{Probability of a Single Cell Destruction}

The average lifetime of a membrane in the external electric field can be
estimated with the help of the following equation (Weaver \& Chismadzhev,
1996):

\begin{equation}
\tau (T,u)=\tau _{\infty }\exp (\Delta F^{\ast }/(1+u^{2}C^{\ast })).
\label{e16}
\end{equation}

Lebedeva (1987) has presented the following estimates for the general lipid
membranes: $\tau _{\infty }\approxeq 0.37$x$10^{-6}$ s , $\omega \approxeq
1.69$x$10^{-11}$ N, $\gamma \approxeq 2$x$10^{-3}$ N m$^{-2}$, $%
C_{m}\approxeq 3.5$x$10^{-3}$ F m$^{-2}$ at $T=298$ K.

Then the breakage probability for a membrane (as a whole cell) during the
impact period of an impulse with duration of $t_{i}$  may be estimated as

\begin{equation}
g=1-\exp (-t_{i}/\tau (T,u)).  \label{e17}
\end{equation}

Taking Eq. (\ref{e16}) into account, we can rewrite Eq. (\ref{e17}) in the
following convenient dimensionless form

\begin{equation}
g(u^{\ast })=1-\exp \left( -\ln 2/\exp a([1-(1-u^{\ast }{}^{2})/(a\Delta u\ln
2/u_{o})]^{-1}-1)\right) ,  \label{e18}
\end{equation}
where $u^{\ast }=u/u_{o}$, $u_{o}=\sqrt{(\Delta F^{\ast }/a-1)/C^{\ast }}$, $%
\Delta u=u_{o}\left( (1-a/\Delta F^{\ast })a\ln 2\right) ^{-1}$ , $a=\ln (t_{i}/(\tau
_{\infty }\ln 2))$, and $du\geq du_{c}=1/(a\ln 2)$ .

We see that $g(u^{\ast })$ is a kind of probability transition function and $u=u_{o}$
corresponds to the midpoint, where $g(u)=1/2$ and $du$ is
the width of this function. We can estimate the following
actual values for the general lipid membranes using the experimental data of
Lebedeva (1987): $u_{o}\approxeq 0.92$V, $\Delta u\approxeq 0.41$, $a\approxeq
3.66 $ and $\Delta u_{c}\approxeq 0.394$ (at $t_{i}=10$ $\mu $s) and $u_{o}\approxeq
0.71$V, $\Delta u\approxeq 0.26$, $a\approxeq 5.97$ and $\Delta u_{c}\approxeq 0.242$
(at $t_{i}=100$ $\mu $s).

\subsection{Description of the Simulation Model}

To achieve the best performance, we have used here a hybrid of MF and
cellular automaton strategies in order to reduce the computational
complexity. First we construct an array of undestroyed cells located in the
sites of a simple cubic lattice. The system has the $N_{x}$x$N_{y}$x$N_{z}$
dimension and periodical boundary conditions are applied in all directions
(Fig. \ref{f3}). We use $N_{x}=N_{y}=10$ and $N_{z}=1000$ in our simulation.
The external electric field of the $E$ strength is applied along the $z$-axis
and the total sample width is $L=d_{c}N_{z}$. The site with intact cell is
marked as occupied. We consider the idealised square pulse sequence with the
pulse duration $t_{i}$, and the pulse repetition time $\Delta t$.

The simulation procedure is done at each time step according to the
following scheme:

a) choose of the next occupied site within a lattice using the linear
search procedure;

b) determination of the number of occupied sites ($n$) among all of its six
near-neighbours, determination of $P_{n}$, and calculation of the
transmembrane voltage $u$ at a given cell with the help of Eqs. (\ref{e01}),
(\ref{e06}) and (\ref{e15});

c) calculation of the probability of a given cell destruction using Eq. (\ref{e18});

d) reiteration of the step (a) until all the occupied lattice sites appear
to be tested.

The total destruction degree $P$ and values of $\sigma _{e}$ and $<$ $%
f_{n}(P,P_{n})>$ are calculated after each pulse. These time steps run until
$P$ reaches its asymptotic value. This model is a simple cellular automaton
for the purposes of simulation of the heterogeneous material breakage kinetics.
Indeed, at the step (b) we test the near-neighbour interior of a given cell
and then we calculate the probability of destruction.

The elevation of the temperature after each pulse was estimated with account
to the Joule heating of material. The ohmic temperature increase $\Delta
T(t_{i})$ after each pulse of $t_{i}$ duration was calculated using the
following equation

\begin{equation}
\Delta T(t_{i})=\frac{d\Delta T}{dN}=\frac{E^{2}\sigma _{e}t_{i}}{C\rho }.
\label{e20}
\end{equation}

We use in all calculations: $d_{c}=10^{-5}$ m, $ d_{m}=10^{-8}$ m,
$\sigma _{c}=0.1-1$ S m$^{-1}$, $\sigma _{m}=10^{-4}-10^{-6} $ S m$^{-1}$,
$t_{i}=10-100$ $\mu $s and putting $C=3.93$ kJ kg$^{-1}$K$^{-1}$ and
$\rho =0.81$x$10^{3} $kg m$^{-3}$, which values are characteristic
for apples at $25{{}^{\circ }}$C (Losano,Urbicain \& Rotstein,1979).

\section{Materials and Methods}

\subsection{Materials}

Freshly harvested apples of Golden Delicious variety were selected for
investigation and stored at $4{{}^{\circ }}$C until required.
In all the cases $W$ was within 80-85\%. We estimated the cell
destruction degree as a ratio of effective conductivities measured before
and after pulsed electric field treatment. The specific conductivities of
samples, both initial (before treatment) and final (after the full treatment
at highest voltages) were within $\sigma _{b}=0.004-0.008$ S m$^{-1}$ and
$\sigma _{a}=0.1-0.2$ S m$^{-1}$, respectively. The specific conductivity of
the apple juice extracted from the sample apples was within $\sigma
_{j}=0.1-0.3$ S m$^{-1}$.

\subsection{Methods}

Thin slices ($6\pm 0.2$ mm thickness and $45\pm 0.5$ mm diameter) were cut
from an apple pap. The conductivity was measured with LCR Meter HP $4284$A
(Hewlett Packard) for the thin apple slice samples and with Conductimetre
HI$8820$N(Hanna Instruments, Portugal) for the apple juice samples at frequency
$50$Hz (this frequency was selected as an optimal in order to remove the
influence of the polarising effects on electrodes and inside the samples).
Figure \ref{f4} is a schematic representation of the experimental
pulsed electric field treatment set-up. The temperature was recorded by the
thermocouple THERMOCOAX type 2 (AB 25 NN, $\pm $0.1${^{\circ }}$C). High
voltage pulse generator, $1500$V-$15$A (Service Electronique UTC,
France) allowed to vary $t_{i}$ within the interval of $10-1000$ $\mu $s (to
precision $\pm 2$ $\mu $s), $\Delta t$ within the interval of $1-100$ ms (to precision
$\pm 0.1$ms) and $N$ within the interval of $1-100000$. All the experiments
were repeated, at least, five times. Pulse protocols and all the output data
(current, voltage, impedance, and temperature) were controlled with the data
logger via Windows $95$ software.

\section{Results and discussion}

Figure \ref{f5} presents the experimental curves of relative conductivity
$\sigma _{a}/\sigma _{b}$ versus number of pulses $N$ for different
values of the electric field strength $E$ with $t_{i}=100$ $\mu $s and $\Delta
t=10$ms. We have obtained practically similar results for $\sigma
_{a}/\sigma _{b}$ vs $N$ dependencies at different values of $\Delta t$
within the interval of $1-100$ms. Hence, pulse repetition time does not
influence our data and we have used $\Delta t=10$ms in all the experiments.
We have observed similar results for $\sigma _{a}/\sigma _{b}$ versus
$Nt_{i}$ (equivalent time of electrical treatment) on $t_{i}$ variation. The
kinetics of temperature evolution $\Delta T$ against $N$ is presented in
Fig. \ref{f6}.

We can conclude out of the data obtained that there exist, at least, two
different stages of the material breakdown evolution. We have divided these
stages conditionally by a horizontal dashed line in the Fig. \ref{f5}. At the
first stage, the sequential and the correlated breakdowns of cells develop
in the system. The time of the first stage changes drastically with $E$
increase. At low values of $E$ we observe a very slow evolution of the
material breakdown. The second stage of material breakdown (over the
horizontal dashed line) flows more rapidly until terminated by an abrupt
total dielectric breakdown of the material. At this moment we observe the
overflow value of the out-of-limit current ($15$A) and stop the further
treatment of material. We observe for each $E$ a certain value of $N_{b}$,
which corresponds to the number of pulses at the moment of the total
dielectric breakdown of material. Figure \ref{f7}a presents $N_{b}$ versus
$E$ (electric field strength) dependence.

The insert in Fig. \ref{f6} shows that practically in all the cases we
observe approximately linear $T$ versus $N$ dependencies. Some deviation
from this linear behaviour can be seen only for small $E$ values
($E=0.2$ kV cm${}^{-1}$)
and at large $N>10^{4}$; we can explain it by the heat exchange with the
outside media. The slopes of the near-linear $T$ versus $N$ curves
correspond to the mean values of $\Delta T(t_{i})=d\Delta T/dN$ averaged
over the total interval of $N$. They are shown in Fig. \ref{f7}b for the
different values of $E$. Here the solid line corresponds to the square-law
$\Delta T(t_{i})$ versus $E$ increase in accordance with relation

\begin{equation}
\Delta T(t_{i})=a+bE+cE^{2},  \label{e21}
\end{equation}
which is consistent with Eq. (\ref{e20}). Here $a=1.116$x$10^{-4}$,
$b=-1.118$x$10^{-5}$, $c=4.9463$x$10^{-8}$ are the values obtained from
root mean square
treatment
of the experimental data for the interval $E<1$ kV cm${}^{-1}$. At large values of
field strength, $E>1$ kV cm${}^{-1}$, we observed a considerable deviation from the
parabolic law of Eq. (\ref{e20}) in the $d\Delta T/dN$ versus $E$ dependency.
The reasons of such deviation are still unclear.

The computer simulation results for the breakage degree ($P$) versus number
of pulses ($N$) and effective conductivity ($\sigma _{e}$) versus number of
pulses ($N$) dependencies at the different values of $E$, $\lambda =0.1$ and
$\lambda =0.01$ are presented at Fig. \ref{f8}a,b. We observe the obvious
step-like behaviour of these $P(N)$ and $\sigma _{e}(N)$ curves describing
the breakage kinetics. This behaviour reflects the ``jamming'' effects
present in the systems under investigation and has a pure geometric origin.

The ``jamming'' problem is very important and occurs in a number of
situations, such as irreversible surface deposition of extended objects,
random sequential adsorption (RSA problem), polymer physics problem or
car-parking problem (J. W. Evans,1993; Nielaba, Privman \& Wang, 1993). The
basic characteristic of this problem is the ``jamming'' coverage that
depends on the type of the lattice and system dimensionality. In the case of
deposition of the definite size particles the ``jamming'' limit is reached when
it is impossible to place any further objects without overlapping the
deposited before. The ``jamming'' effects in the dielectric breakdown
problem have the following origin: when a cell gets punctured, there appear
the bridging effects, which cause the abrupt fall of destruction probability
for cells surrounding the punctured one. In the frame of our model, this
bridging effect results from deviations of the local conductivity near the
selected cell $n$ (see, Eq. (\ref{e14})) against the average effective
conductivity of the whole media $\sigma _{e}$.

In the RSA problem, the cell puncture is equivalent to the lattice site
occupation, thus, occupation of definite site results in abrupt fall of the
occupation probability for sites neighbouring the punctured one. The
dielectric breakdown problem in such formulation is very similar to the RSA
problem, except for successive ''jamming'' limits, which correspond to the site
occupation with one, two, and et-cetera near-neighbours. We have carried out
the Monte-Carlo simulation for the random sequential occu$^{{}}$pation of
sites in the simple cubic lattice with j neighbour punctured cells in order
to find the ``jamming'' limits $P(j)$. We have obtained the following
values for the lattice of 200x200x200 dimension by averaging the calculation
results over the 10 different initial configurations: $P(0)=0.305047\ldots $%
, $P(1)=0.42026\ldots $, $P(2)=0.49368\ldots $, $P(3)=0.544401\ldots $, $%
P(4)=0.59321\ldots $, $P(5)=0.64347\ldots $, and $P(6)=1$.

The ``jamming'' effects in the dielectric breakdown problem result in
existence of the saturation regimes in the breakdown kinetics. The first
``jamming'' step is clearly observed in the Fig. \ref{f8} for the small
values of $E/u_{o}<$1 at $P=P(0)\approx 0.3$. The next ``jamming'' steps
are less pronounced and can be clearly observed only at small values of $\Delta u$%
. The initial increase on the $P$ versus $N$ curves corresponds to the
breakdown development without any near-neighbours bridging effects. At large
values of $E/u_{o}>1.5$ we observe the jump on the next ``jamming'' steps ($%
P(1)$, $P(2)$) with the subsequent saturation regime.

We have observed the similar saturation regimes in experimental observations
of the kinetics of dielectric breakage in thin apple slices (see
Fig. \ref{f5}), particularly, at small values of $E<0.5$ kV cm$^{-1}$.
The step-like behaviour is not so pronounced in experimental results as in computer
simulation data presented in the Fig. \ref{f8}; this difference results from
the simulation model restrictions and, partially, from the lattice nature.

Figure \ref{f8}c presents the curves of the calculated breakage degree $P$
versus relative ohmic temperature increase $\Delta T({{}^{\circ }}$C$)$ at
different values of the electric strength $E$ for $\lambda =0.1$ and $%
\lambda =0.01$. These calculations were done using Eq. (\ref{e20}).
These data allow us to understand the
correlation existing between the achieved breakdown degree and the power
consumption, which is proportional to $\Delta T$.

When the electric strength ($E$) values are low, the high destruction degree
($P$) can be achieved only on account of high power consumption and,
correspondingly, with high overheat of the surrounding medium. However, the
weakness of the electric field  treatment is the possibility of a
``jamming'' regime, when the increase in power consumption fails to result in
$P$ increase, and poor control over the course of process. As far as it is
difficult to choose precisely the required electric treatment mode for the
given field strength values, here the transition to the mode of the
overflowing out-of-limit current value
occurs readily. In this case we
stop the further treatment of a material and actually don't achieve the high
values of $P$.

Figure \ref{f9} presents the curves of the number of pulses in the total
breakdown point ($N_{b}$) versus reduced electric strength ($E/u_{o}$) for
different values of $\Delta u/u_{o}$ and $\lambda $ . We see that the character of
$N_{b}$ vs $E$ curves depends on the values of $\Delta u/u_{o}$ and $\lambda $. The
$N_{b}$ increases with decrease of $\Delta u/u_{o}$ at given $E$ value. Unfortunately, we are
unable to make more strict comparison between theoretical and experimental
data, as far as we have no precise data for $u_{o}$, $\Delta u$ and $\lambda $
parameters present in computer model. However, the general character of
calculated $N_{b}$ vs $E$ dependencies (Fig. \ref{f9}) correlates with
experimental data presented in Fig. \ref{f7}a.

\section{Conclusion}

The simplified dielectric breakage model based on effective media
approximation is proposed. This approximation includes equations with the
same morphology parameters as in percolation theory. The normalised
transmembrane voltage $f=u/d_{c}E$ versus media breakage degree $P$
dependence is obtained; this dependence is useful for estimating the
transmembrane voltage and of the cell breakdown probability under the PEF
treatment of cellular material. The most important feature of the proposed model is
the existence of the ``jamming'' behaviour occurring in experimental
observations of the biological tissue breakage. The different transitions
corresponding to the different ``jamming'' steps are identified. The
experimental study has yielded information about the material destruction
degree and temperature elevation versus the time of PEF
treatment at different values of field strength $E$ in the interval of
$0.2-2.2$ kV cm$^{-1}$. The proposed simulation model gives results
consistent with the general trends observed in experimental breakage
kinetics. The electric treatment is the most efficient at high values of the
electric field strength $E$. However the efficiency of such electric
treatment is limited by manifestations of the ``jamming'' effect, as well as
by the processes of electric treatment upset. Reduction of $E$ values allows
controlling the treatment process more precisely and prevents from
uncontrolled breakdown but it is accompanied with increase of electric power
consumption.

\section*{Acknowledgements}

The authors would like to thank the ``Pole Regional Genie des Procedes``
(Picardie, France) for providing financial support.
Authors also thank Dr. N. S. Pivovarova for help with the
preparation of the manuscript.


\section*{References}


De Arcangelis L., Redner, S., \& Herrmann H. (1985). A random fuse model for
breaking processes. J.Phys.(Paris) Lett, 46, L585.

Barbosa-C\'{a}novas, G.V., Pothakamury, U.R., Palou, E., \& Swanson, B.G.
(1998). Nonthermal Preservation of Foods (pp. 53-72). New York: Marcel
Dekker.

Chang, D.C., Chassy, B.M., Saunders, J.A., \& Sowers, A.E. (1992). Overview
of electroporation and electrofusion. In~D.C. Chang, B.M. Chassy, J.A.
Saunders \& A.E. Sowers (Eds.). Guide to electroporation and electrofusion
(pp. 1-6). San Diego: Academic Press.

Duxbury, P.M. , Beale, P.D. (1995), Breakdown of Two-Phase Random Resistor
Networks. Phys. Rev. B 51, 3476.

Evans, J.W. (1993). Random and Cooperative Sequential Adsorption. Rev. Mod.
Phys., 65, 1281.

Hulsheger, H., Potel, J., \& Niemann, E.G. (1983). Electric field effects on
bacteria and yeast cells. Radiat. Environ. Biophys, 22, 149-162.

Knorr, D., Geulen, M., Grahl, T., \& Sitzmann, W. (1994). Food application
of high electric field pulses. Trends in Food Science \& Technology, 5,
71-75.

Knorr, D. (1997). Impact of High Intensity Electric Field Pulses on Plant
Cells and Tissues. Proceedings of Conference of Food Engineering (p 66f),
Los Angeles, California.

Kotnik, T, Miklavcic, D., \& Slivnik, T. (1998). Time course of
transmembrane voltage induced by time-varing electric fields -- a method for
theoretical analysis and its application. Bioelectrochem. Bioenerg., 45,
3-16.

Landauer, R. (1978). Electrical transport and optical properties of
inhomogeneous media. American Institute of Physics, 40, 2.

Lebedeva, N.E. (1987). Electric Breakdown of Bilayer Lipid Membranes at
Short times of Voltage Effect. Biologicheskiye Membrany, 4 (9), 994-998 (in
Russian).

Lebovka, N.I., \& Mank, V.V.(1992). Phase Diagram and Kinetics of
Inhomogeneous Square Lattice Brittle Fracture. Physica A, 181, 346- 363.

Losano, J.E., Urbicain, M.J., \& Rotstein, E. (1979). Thermal conductivity
of apples as a function of moisture content. J Food Sci, 44, 198-199.

McLachlan, D.S. (1989). The complex permittivity of emulsions: an effective
media-percolation equation. Solid State Communications, 72(8), 831-834.

Neumann, E., Sprafke, A., Boldt, E., \& Wolf, H. (1992). Biophysical
Considerations of Membrane Electroporation. In D.C. Chang, B.M. Chassy, J.A.
Saunders \& A.E. Sowers (Eds.). Guide to electroporation and electrofusion
(pp. 77-117). San Diego: Academic Press.

Nielaba, P., Privman, V., \& Wang, J.-S. (1993). Irreversible multilayer
adsorption. In D.P. Landau, K.K. Mon, H.-B. Sch\"{u}ttler (Eds.). Computer
Simulation Studies in Condensed-Matter Physics VI (p.143). Springer
Proceedings in Physics, vol. 76, Heidelberg-Berlin: Springer-Verlag.

Press, W.H., Teukolsky, S.A., Vetterling, W.T., \& Brian, P. (1997).
Flannery Numerical Recipes in Fortran 77: The Art of Scientific Computing
(vol.1). Cambridge: Cambridge University Press.

Sahimi, M. (1994). Applications of Percolation Theory. London: Taylor and
Francis.

Scheglov, Y.A. (1983). Electroplasmolysis: A new method for treating fruits
and vegetables. J. Panorama. Licensintorg, 9, 30.

Weaver, J.C., \& Chismadzhev, Y.A. (1996). Theory of electroporation: a
review. Bioelectrochem. Bioenerg., 41, 135-160.

Zimmermann, U., Pilwat, G., \& Riemann, F. (1974). Dielectric breakdown of
cell membranes. Biophys. J., 14, 881-889.


\begin{figure}[tbp]
\caption{Normalised transmembrane voltage $f=u/d_{c}E$ versus media
destruction degree $P$ dependencies calculated for the MF approximation
using Eqs. (\ref{e02}),(\ref{e06}) (curve 1), for the LC approximation using
Eqs. (\ref{e06}), (\ref{e12}) (curve 2) and for the RSO model using results of
simulation (curve 3). The dashed curve 4 corresponds to $f$ dispersion $%
\Delta f$ in RSO model. The calculations were performed at $t=2.3$, $%
P_{p}=0.307$, $d_{c}=10^{-5}$ m, $d_{m}=10^{-8}$ m, $\protect\sigma _{c}=
1$ S m$^{-1}$, $\protect\sigma _{m}=10^{-5}$ S m$^{-1}$, and
$\protect\lambda =0.01$.}
\label{f1}
\end{figure}

\begin{figure}[tbp]
\caption{Normalised transmembrane voltage $u/d_{c}E$ versus media destruction
degree $P$ at various $n$ (the number of intact cells neighbouring a
given cell). The dashed line corresponds to the example of $u/d_{c}E$ behaviour
with increase of the number of destroyed near-neighbours in the course of
the media breakage. The calculations were performed at $t=2.3$,
$P_{p}=0.307$, $d_{c}=10^{-5}$ m, $d_{m}=10^{-8}$ m, $\protect\sigma _{c}=1$
S m$^{-1}$, $\protect\sigma _{m}=10^{-5}$ S m$^{-1}$, and
$\protect\lambda =0.01$ for cells arranged in the sites
of a simple cubic lattice.}
\label{f2}
\end{figure}

\begin{figure}[tbp]
\caption{The model of the cellular material structure used in the computer
simulation.}
\label{f3}
\end{figure}

\begin{figure}[tbp]
\caption{Schematic representation of the experimental set-up used in the
study of pulsed electric field treatment of the apple slices.}
\label{f4}
\end{figure}

\begin{figure}[tbp]
\caption{Relative conductivity $\protect\sigma _{a}/\protect\sigma _{b}$
versus number of pulses $N$ at different values of the electric field
strength $E$, $t_{i}=100 \protect$ $\mu $s and $\Delta t=10$ ms for thin apple
slices at $25{{}^{\circ}}$C.}
\label{f5}
\end{figure}

\begin{figure}[tbp]
\caption{Relative ohmic temperature increase $\Delta T$(${{}^{\circ }}$C)
versus number of pulses $N$ at different values of the electric field
strength $E$, $t_{i}=100 \protect$ $\mu $s and $\Delta t=10$ ms for thin apple
slices at 25${{}^{\circ}}$C. Insert shows the same data for small $N$
values in linear co-ordinates $T$ vs $N$.}
\label{f6}
\end{figure}

\begin{figure}[tbp]
\caption{Dependencies of the number of pulses in the total breakdown point
$N_{b}$ and of the mean ohmic temperature increase $\Delta T(t_{i})$ after
each pulse of $t_{i}$ duration against the electric strength $E$ at $%
t_{i}=100 \protect$ $\mu $s and $\Delta t=10$ ms for thin apple slices at $25{%
{}^{\circ }} $C.}
\label{f7}
\end{figure}

\begin{figure}[tbp]
\caption{Calculated breakage degree $P$, effective conductivity $\protect%
\sigma _{e}$ versus number of pulses $N$ dependencies (a,b) and breakage
degree $P$ versus relative ohmic temperature increase $\Delta $T (${%
{}^{\circ }}$C) dependencies (c) at different values of electric field strength $E$
for $\protect\lambda =0.1$ and $\protect\lambda =0.01$. The calculation are
performed at $t=2.3$, $P_{p}=0.307$, $\Delta u=0.35$, $u_{o}=1$, $t_{i}=100 \protect%
$ $\mu $s, $d_{c}=10^{-5}$ m, $d_{m}=10^{-8}$ m, $\protect\sigma _{c}=1$ S m$^{-1}$%
, $\protect\sigma _{m}=10^{-5}$ S m$^{-1}$ ( $\protect\lambda =0.01$), and $%
\protect\sigma _{c}=0.5$\ S m$^{-1}$, $\protect\sigma _{m}=5$x$10^{-5}$ S m$%
^{-1}$ ($\protect\lambda =0.1$).}
\label{f8}
\end{figure}

\begin{figure}[tbp]
\caption{Calculated number of pulses in total breakdown point $N_{b}$ versus
reduced electric strength $E/u_{o}$ for different values of $\Delta u/u_{o}$ and $%
\protect\lambda $. The calculation are performed at $t=2.3$, $P_{p}=0.307$, $%
\Delta u=0.35$, $u_{o}=1$, $t_{i}=100\protect$ $\mu $s, $d_{c}=10^{-5}$ m, $%
d_{m}=10^{-8}$ m, $\protect\sigma _{c}=1$ S m$^{-1}$, $\protect\sigma %
_{m}=10^{-5}$ S m$^{-1}$ ( $\protect\lambda =0.01$), and $\protect\sigma %
_{c}=0.5$\ S m$^{-1}$, $\protect\sigma _{m}=5$x$10^{-5}$ S m$^{-1}$ ($%
\protect\lambda =0.1$).}
\label{f9}
\end{figure}

\end{document}